\documentclass[reprint,amsmath,amssymb,aps,prresearch,noeprint,longbibliography]{revtex4-2}
\usepackage[utf8]{inputenc}
\usepackage{graphicx}
\usepackage{amsmath}
\usepackage{xspace}
\usepackage{chemformula}

\newcommand{\py}{Py\xspace}  

\begin{document}

\title{Barkhausen noise from formation of 360$^{\circ}$ domain walls\\ 
in disordered permalloy thin films}
\author{Sami Kaappa}
\author{Lasse Laurson}
\affiliation{Computational Physics Laboratory, Tampere University, P.O. Box 692, FI-33014 Tampere, Finland}

\date{\today}

\newlength{\figwidth}
\setlength{\figwidth}{0.95\columnwidth}
\newlength{\widefig}
\setlength{\widefig}{0.9\textwidth}

\begin{abstract}
Barkhausen noise in disordered ferromagnets is typically understood to originate primarily from jerky field-driven motion of domain walls. We study the magnetization reversal process in disordered permalloy thin films using micromagnetic simulations, and find that the magnetization reversal process consists of gradual formation of immobile 360$^{\circ}$ domain walls via a sequence of localized magnetization rotation events. The density of 360$^{\circ}$ domain walls formed within the sample as well as the statistical properties of the Barkhausen jumps are controlled by the disorder strength.     
\end{abstract}

\maketitle

\section{Introduction}
Permalloy (Ni$_{80}$Fe$_{20}$, \py) thin films are important ferromagnetic systems in spintronics and magnonics applications~\cite{Gardelis1999, Jedema2001, Miao2013, Haidar2019, Bommanaboyena2021} due to their soft magnetic response and low fabrication cost. \py thin films exhibit a hysteretic response to external fields and display crackling noise~\cite{Roy2020_1, Roy2020_2, Yang2005} known as Barkhausen noise (BN)~\cite{Durin2004}. The statistical properties of BN are typically found to obey power law behaviour~\cite{Sethna2001,Durin2004,zapperi1998dynamics}. Statistical physicists often model BN by employing either so-called front propagation or nucleation models, such as the random-field Ising model (RFIM)~\cite{Puppin2000, Kim2003, Zani2004, Yang2005, Santi2006, Yang2006, Janicevic2018, Roy2020_1, Roy2020_2,mughal2010effect}, with the BN in the neighborhood of the coercive field (where the largest BN amplitude is found) understood to originate from the jerky field-driven motion of domain walls (DWs). However, given the vanishing magneto-crystalline anisotropy of \py, Ising-type models which inherently assume a strong uniaxial anisotropy are not suitable to properly characterize its magnetization reversal process.

In this work, we employ full micromagnetic simulations of disordered \py thin films to properly capture the details of the field-driven magnetization reversal process and the related statistical properties, including the emergence of BN. Unlike simple Ising-type models, micromagnetic simulations provide a proper description of the full vectorial nature of the order parameter (magnetization) and include all the relevant energy terms (such as the demagnetizing energy). Contrary to the paradigm of BN due to jerky motion of DWs, we find that field-driven BN in \py thin films is due to the gradual formation of largely {\it immobile} 360$^{\circ}$ DWs~\cite{Smith1962,Cho1999,Rippard2000, Castano2003, Hehn2008, Zhang2016, Liedke2006, OShea2015, Chowdhury2018,Muratov2008, Dean2011, Su2017, Li2021} via a sequence of localized and intermittent magnetization rotation events. Once formed, the 360$^{\circ}$ DWs get progressively narrower due to an increasing driving field induced effective anisotropy, before disappearing as the oppositely saturated state is reached. Both the emerging configuration of 360$^{\circ}$ DWs (e.g., their density) and the statistics of BN are found to be dependent on disorder strength, suggesting that the magnetization reversal process in disordered \py thin films is governed by disorder-induced criticality~\cite{Sethna2001}, but with an important difference compared to Ising models: no infinite, spanning avalanche (i.e., an avalanche spanning the finite system along at least one of its dimensions, resulting in a jump of magnetization in an infinite system below the critical disorder) is observed for weak, subcritical disorder. The present work therefore focuses on investigating how the Barkhausen effect and the related jump size distributions emerge from the previously unidentified mechanism of gradual formation of immobile 360$^\circ$ domain walls in thin films with negligible magnetocrystalline anisotropy, as opposed to the well-known mechanism related to the jerky motion of domain walls.

\begin{figure}[t!]
    \centering
    \includegraphics[width=\figwidth]{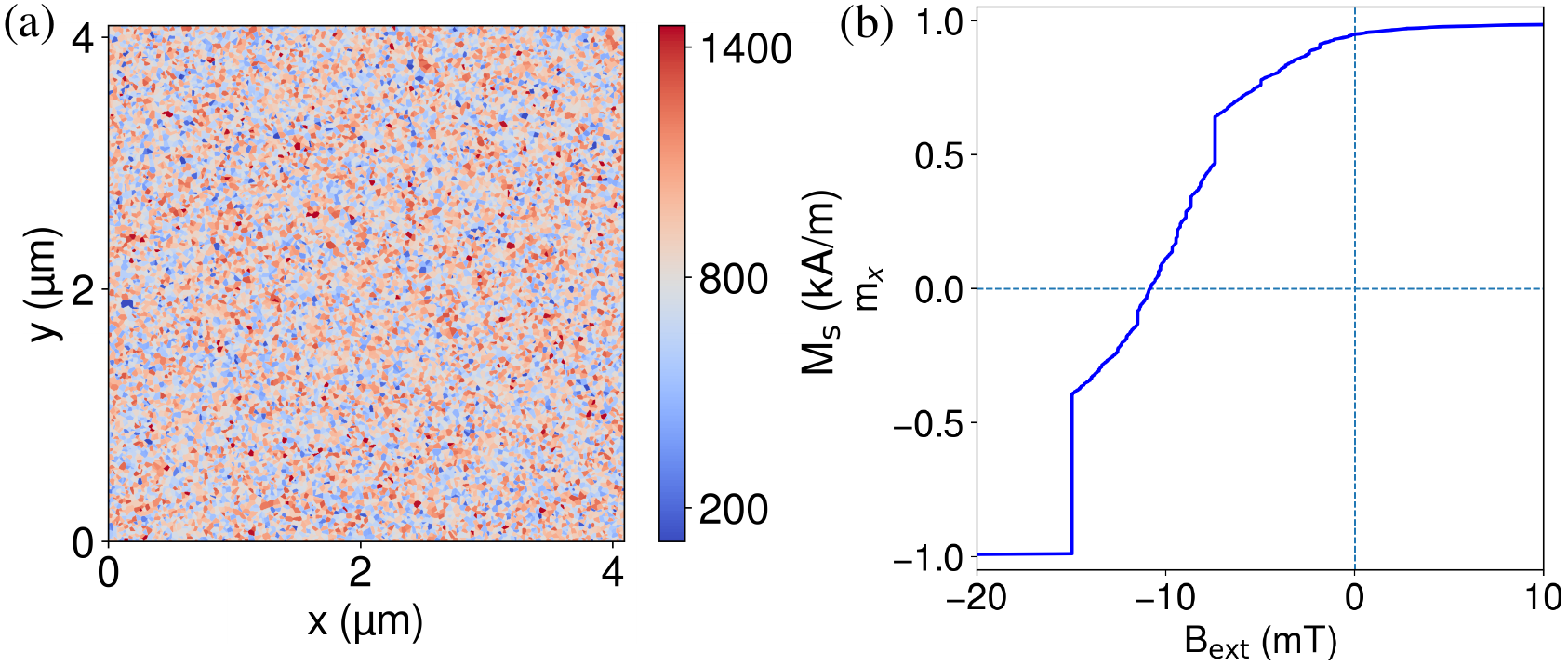}
    \caption{(a) Example of the spatial variation of $M_s$, sampled from the normal distribution with mean 800 kA/m and standard deviation of 30 \% of the mean value. (b) Example evolution of $m_x$ with quasistatically decreasing $B_\mathrm{ext}$.}
    \label{fig:system}
\end{figure}

\section{Computational methods}
Our micromagnetic simulations are performed using the GPU-accelerated simulation code mumax3~\cite{Vansteenkiste2014}. Typically, micromagnetic simulations rely on the dynamical approach where the magnetization dynamics is governed by the Landau-Lifshitz-Gilbert equation, written as
\begin{equation}\label{eq:LLG}
    \frac{\partial \vec M}{\partial t} = \frac{\gamma}{1 + \alpha^2} \left(\vec M \times \vec H_\mathrm {eff} + \alpha \vec M \times \left(\vec M \times \vec H_\mathrm {eff}\right)\right),
\end{equation}
where $\vec M=M_s\hat{m}$ is the magnetization vector, $M_s$ is the saturation magnetization, $\hat m = m_x \hat x + m_y \hat y + m_z \hat z$ is the unit vector pointing along the magnetization, $\gamma$ is the gyromagnetic ratio of electron, $\alpha$ is the Gilbert damping constant, and $\vec H_\mathrm {eff}$ is the effective field. The total energy of the system (and hence $\vec H_\mathrm {eff}$) consists of contributions from exchange, demagnetization, and Zeeman terms as described in detail in the Supplemental Material~\cite{SM}. Due to the vanishing magnetocrystalline anisotropy of \py, the anisotropy energy is neglected here, that is the magnetocrystalline anisotropy constants are set to zero.

\begin{figure*}[t!]
    \centering
    \includegraphics[width=\widefig]{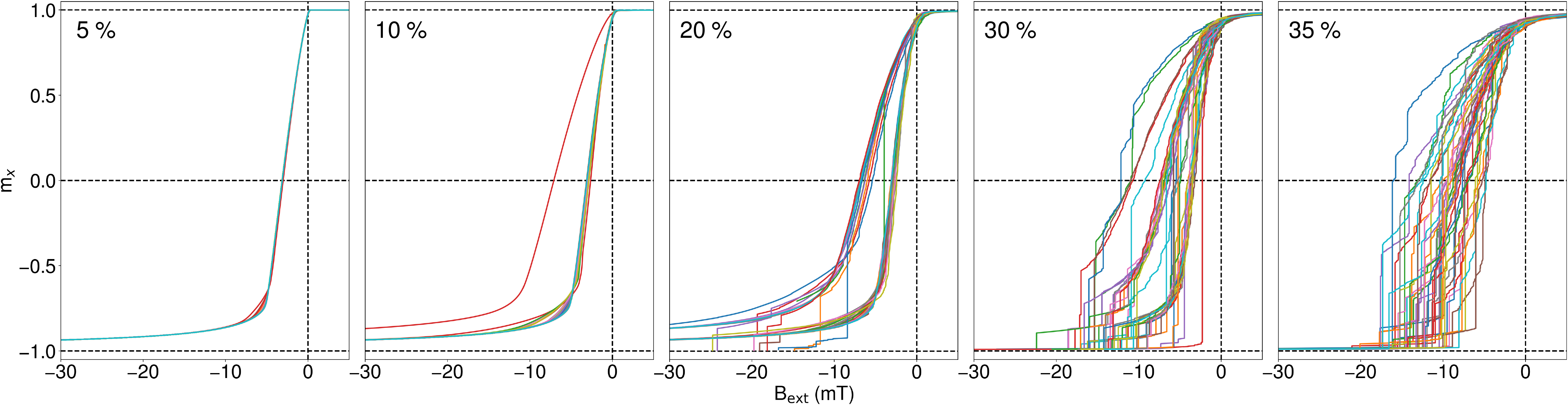}
    \caption{Hysteresis curves for $m_x(B_\mathrm{ext})$ of all simulations. The external field $B_\mathrm{ext}$ is decreased from 100 mT to $-100$ mT. The disorder strength is indicated in the upper left corner of each panel.}
    \label{fig:multihysteresis}
\end{figure*}

An example of a simulated system is shown in Fig. \ref{fig:system}(a). We discretize the square thin film on a finite difference grid with spacing of 4 nm in the $x$ and $y$ in-plane directions where the linear system size is 4096 nm and consider a single 16 nm long cell in the perpendicular $z$ direction (the film thickness). Periodic boundary conditions are employed in the in-plane $x$ and $y$ directions. We use the common literature values for the exchange stiffness $A = 13$ pJ/m and the average saturation magnetization $M_s = $ 800 kA/m. Structural disorder is introduced via a spatially varying $M_s$ by performing a 2D Voronoi tessellation on each sample with the grain size of 30 nm (the disorder correlation length), and assigning a random value of $M_s$ in each grain from a normal distribution with mean of 800 kA/m and standard deviations (the disorder strengths) between 5 \% and 35 \% of the mean value (negative $M_s$ values are avoided by re-drawing until a positive value is obtained) ~\cite{Min2010, Leliaert2014}. This way, we establish a random component in the energy landscape of the magnetization state, characterized by the correlation length (Voronoi grain size) and magnitude (disorder strength) of the disorder. In a material, such disorder can be caused by inhomogeneous distribution of elements in the sample, impurity material, defects, and other lattice imperfections. Introducing the disorder by parameter variation in the zero-anisotropy case can be done through variation of either $M_s$ or $A$ or both. Here we investigate the effect of the $M_s$ variation only. According to our test runs, simultaneous variation of exchange stiffness $A$ does not alter the magnetization dynamics qualitatively but the dynamical process is dominated by the spatial $M_s$ variation. 

In each run, the magnitude $B_\mathrm{ext}$ of the external field $\vec B_\textrm{ext}=B_\textrm{ext}\hat{x}$ is swept from 100 mT (at which point the system is close to saturation with $m_x \approx 1$) to $-$100 mT. The run is considered finished once $m_x$ falls below $-0.98$ or the external field reaches $-100$ mT. This procedure thus produces half of the hysteresis loop where $m_x$ reverses from a value close to $+1$ to a value close to $-1$.

The simulation times of dynamic simulations are in practice limited to the microsecond range, hence seriously limiting the range of accessible field frequencies. Therefore, we mainly focus on {\it quasistatic} simulations, where the total energy of the system is consecutively minimized while altering $B_\textrm{ext}$ between each minimization in small steps $\Delta B_\textrm{ext}$. To validate the use of quasistatic simulations, we present in  Fig. S1 (Supplemental Material~\cite{SM}) results for both dynamical and quasistatic simulations, showing how the dynamic hysteresis loops approach the hysteresis loop obtained from the quasistatic simulations in the low frequency limit of the sinusoidal driving field. In what follows, we thus consider quasistatic simulations with $\left|\Delta B_\textrm{ext}\right| = 4$~\textmu T; see Fig. S2 \cite{SM} for the justification of the selected value. For an example of the half-loop produced by this protocol, see Fig.~\ref{fig:system}(b). For each disorder strength, we collect statistics by running 40 simulation runs with different disorder realizations.

\section{Results and discussion}
\subsection{Hysteresis curves}
Figure \ref{fig:multihysteresis} shows all of the simulated hysteresis curves for each disorder strength. The curves for 5 and 10 \% disorders are quite smooth, i.e., Barkhausen jumps are largely absent. The magnetization is being rotated in a smooth and continuous fashion as the applied field is driven from 100 mT to $-100$ mT, i.e., no infinite spanning avalanche takes place unlike in RFIM for weak disorder. Note also that the magnetization does not saturate to $-1$ before $B_\textrm{ext}=-30$ mT. In contrast, the systems with the highest disorder strengths, 30 \% and 35 \%, exhibit bursty behaviour with more frequent and larger abrupt avalanches. The saturated state with $m_x = -1$ is reached before $B_\textrm{ext}=-30$ mT for most runs. The curves with 20 \% disorder show features of both kinds: long smooth sections interrupted by occasional abrupt jumps.

Following this observation, let us make a distinction between the parts of the hysteresis curve where (1) the magnetization is rotated smoothly and continuously and (2) a large part of the magnetization is changed abruptly (Barkhausen jumps). We note that the tangential slopes of the smooth parts in between avalanches seem to have a characteristic magnitude. Also, the curves with 10 and 20 \% disorder strengths seem to be divided into two separate subgroups, as well as the 30 \% curves into four subgroups (Fig.~\ref{fig:multihysteresis}). In the following, we will connect all of these features of the hysteresis curves to the disorder strength, the respective Barkhausen jump size distributions, and the magnetization reversal mechanisms, including the appearance of the 360$^{\circ}$ DWs.

\begin{figure}[t!]
    \centering
    \includegraphics[width=\figwidth]{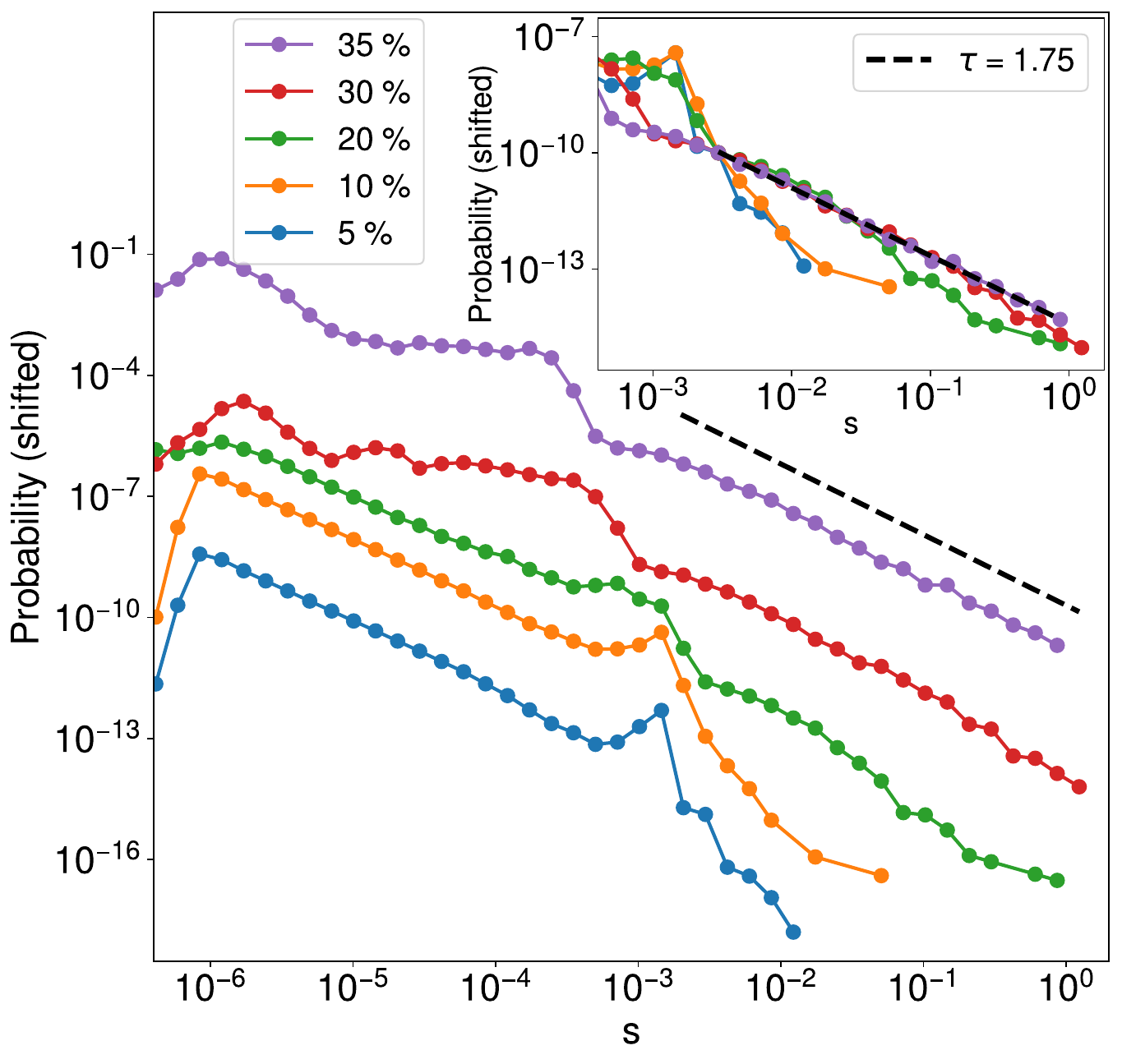}
    \caption{Distributions (shifted vertically for clarity) of the Barkhausen jump sizes $s=|\Delta m_x|$ for different disorder strengths. The exponent $\tau=1.75 \pm 0.02$ is determined by fitting a power law to the tail of the 35 \% distribution. The inset includes the same data stacked on top of each other to highlight the dependence of $s_0$ on the disorder strength.}
    \label{fig:distributions2}
\end{figure}

\subsection{Barkhausen jump size distributions}
Figure \ref{fig:distributions2} shows the Barkhausen jump size distributions, that is the distributions of the absolute changes $s \equiv |\Delta m_x|$ of $m_x$ during each field step in the quasistatic simulations. Based on the renormalization group theory and previous studies of Barkhausen noise statistics \cite{Durin2004}, we expect that the tails of size distributions $P(s)$ follow truncated power laws with a critical exponent $\tau$,
\begin{equation}\label{eq:powerlaw}
    P(s) = s^{-\tau} g(s/s_0)
\end{equation}
where $g(x)$ is a cutoff scaling function, and $s_0$ a cutoff scale parameter. The tails of the distributions in Fig.~\ref{fig:distributions2} do behave according to Eq.~(\ref{eq:powerlaw}) with a disorder-dependent $s_0$ (see below for more details), but a bump exists in each distribution below the power-law part, introducing a characteristic length scale to the statistics. Increasing the disorder strength leads to translation of the bump to smaller jump sizes and possibly to broadening of the bump. We attribute the positions of the bumps to the largest magnitudes for the tangential slopes in the hysteresis curves where the magnetization reversal is smooth. Avalanches exceeding the characteristic size of the bumps, obeying Eq.~(\ref{eq:powerlaw}), are taken to be the actual Barkhausen jumps. 

The disorder strength of 35 \% seems to be close to the critical disorder strength since the tail of $P(s)$ exhibits a clear power law behaviour. Even though the expected shape of Eq.~(\ref{eq:powerlaw}) includes the cutoff function $g(x)$, we employed a plain power law fit with $g=1$ in the absence of properly resolved cutoffs due to limited statistics. Fitting a power law to the tail of $P(s)$  for 35 \% disorder results in $\tau = 1.75 \pm 0.02$. As illustrated in the inset of Fig.~\ref{fig:distributions2}, the distributions for the other disorder strengths fall off from the power law trend in the order given by the disorder strength, i.e., $s_0$ in Eq.~(\ref{eq:powerlaw}) is disorder-dependent. This trend is expected when approaching the critical disorder strength, and so we conclude that the critical disorder strength here is likely to be close to 35 \%. In the literature, the reported values for the critical exponent $\tau$ for \py thin films are diverse: 1.65 \cite{Roy2020_1}, 1.33 \cite{Yang2005}, 1.45 \cite{Yang2006}, and 1.6 \cite{Wiegmanthesis}. Our estimate is close to those reported in Refs.~\cite{Roy2020_1} and~\cite{Wiegmanthesis}.

Another interesting feature in the distributions is the almost perfect power law behaviour for jump sizes smaller than that of the bumps of the 5-20 \% disorder distributions; this seems to be absent in the stronger disorder statistics. These very small magnetization changes correspond to tangential slopes of $m_x(B_\textrm{ext})$ smaller in magnitude than the ones of the bump, and originate from the parts of the hysteresis curves where the magnetization is close to saturation. It is worth noting that the bump positions depend also somewhat on the field step used in the simulations; see Supplemental Material \cite{SM} and Fig. S2 therein.

\begin{figure*}[t!]
    \centering
    \includegraphics[width=\widefig]{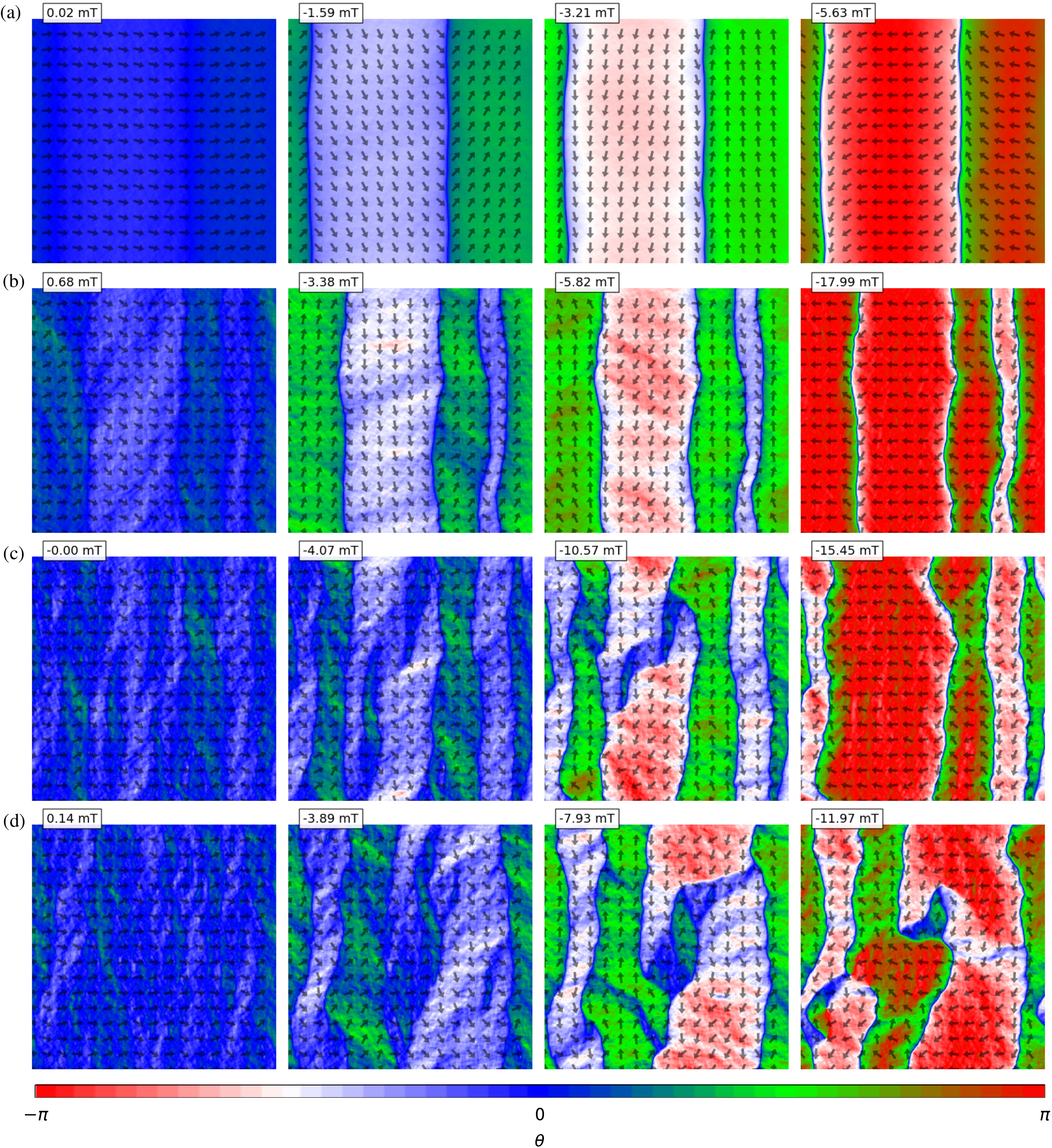}
    \caption{Examples of the magnetization reversal mechanism with (a) 5 \%, (b) 20 \%, (c) 30 \%, and (d) 35 \% disorder strengths. The arrows denote the local magnetization direction in every 64th grid point. The background color indicates the direction of the local magnetization $\vec m$ as given by the color bar; $\theta$ denotes the angle between the $x$-axis and $\vec m$ in radians. The $B_\mathrm{ext}$ values corresponding to each snapshot are indicated in the top left corners.}    \label{fig:mechanism}
\end{figure*}

\subsection{Magnetization reversal processes}
\begin{figure}[h!]
    \centering
    \includegraphics[width=0.8\figwidth]{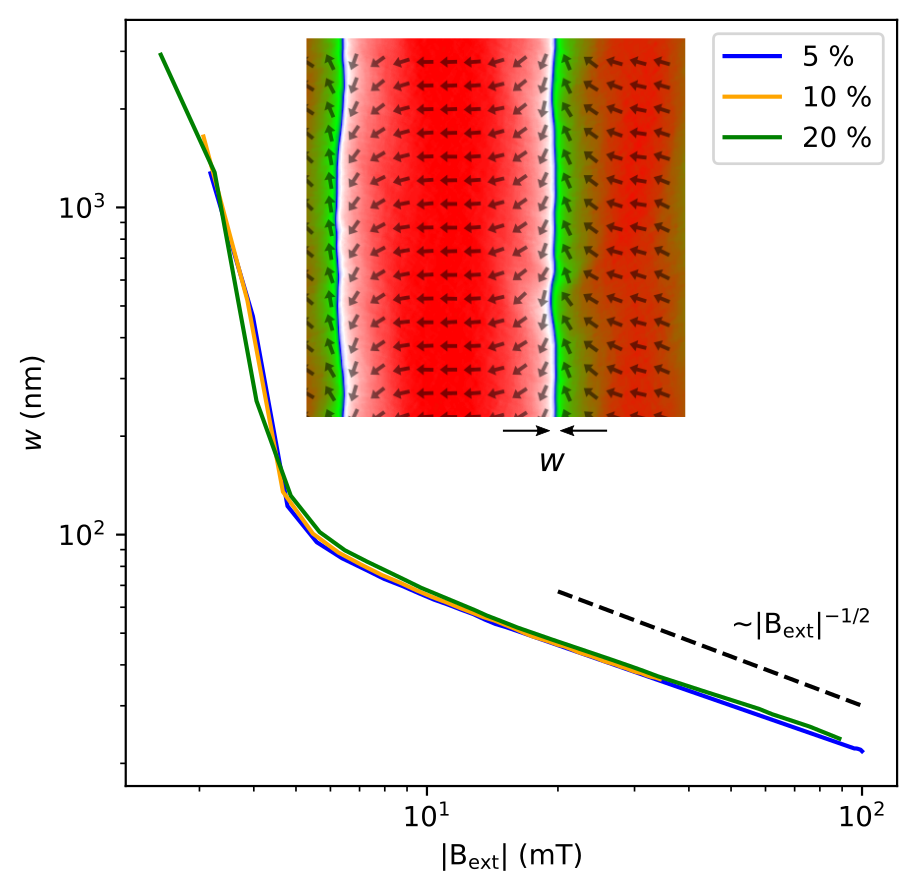}
    \caption{360$^\circ$ DW width $w$ as a function of $|B_\textrm{ext}|$ in selected runs with varying disorder strength where two 360$^\circ$ DWs are formed. $w$ evolves approximately as $|B_\mathrm{ext}|^{-1/2}$ (dashed line). The inset shows an example configuration where $w$ is evaluated at $B_\textrm{ext} =  -5.6 $ mT for 5 \% disorder. $w$ is taken to be the distance between the two points at which the magnetization profile $m_x(x)$, averaged over $y$, crosses 0.}
    \label{fig:dwwidths}
\end{figure}

To understand the origin of the observed subgroups in the hysteresis curves we proceed to study the disorder-dependent magnetization reversal mechanisms. For 5 \% disorder (Fig. \ref{fig:mechanism}(a)), we observe that, interestingly, a structure of two 360$^\circ$ DWs is formed at the late stages of the reversal process. This formation process proceeds via a gradual rotation of the magnetization in opposite directions within two "domains", such that the eventual 360$^\circ$ DWs are formed via 90$^\circ$ DWs at $B_\textrm{ext}=-1.59$ mT and 180$^\circ$ DWs at $B_\textrm{ext}=–3.21$ mT. Due to topological protection~\cite{Zhang2016}, the 360$^\circ$ DWs remain in the system until the end of the simulation, resulting in the absence of full negative saturation of $m_x$ for the $B_\textrm{ext}$-values considered. Upon making $B_\textrm{ext}$ more negative, $m_x$ is slowly approaching $-1$ since the widths $w$ of the DWs decrease with increasing the $B_\textrm{ext}$ (see Fig.~\ref{fig:dwwidths} and the related discussion below). Each of the runs with 5 \% disorder exhibits the formation of two 360$^{\circ}$ DWs, and the hysteresis curves of the different realizations overlap almost perfectly (Fig.~\ref{fig:multihysteresis}).

For 10 and 20 \% disorder, the two subgroups of the hysteresis curves in Fig.~\ref{fig:multihysteresis} correspond to two and four 360$^\circ$ DWs per simulation box, respectively, and some of the 360$^\circ$ DW structures disappear before the applied field reaches $-25$ mT; this is seen as $m_x$ reaching $-1$ in Fig.~\ref{fig:multihysteresis}. The reversal mechanism in an example run where four 360$^\circ$ DWs are formed is presented in Fig.~\ref{fig:mechanism}(b). The mechanism for the formation of the DWs is similar to that of the 5 \% disorder in Fig.~\ref{fig:mechanism}(a), but there is more spatial variation in the local magnetization and the DWs appear more rough. As expected, even more roughness is present in the example cases with 30 and 35 \% disorders shown in Figs.~\ref{fig:mechanism}(c) and (d), where the final magnetization structures are more complicated, and features resembling 360$^\circ$ DWs can be recognized only in parts of the system. The magnetization configurations shown in both Figs. \ref{fig:mechanism}(c) and (d) are fully reversed towards the field direction at the next field step. A closer look at the 30 \% disorder strength hysteresis curves in Fig.~\ref{fig:multihysteresis} reveals that there are distinct subgroups of curves that correspond to varying number of DWs (up to 6 or 8) in the system. Four example runs with 30 \% disorder are highlighted in Fig.~S3 \cite{SM}, further illustrating the different magnetization reversal mechanisms involving formation of different numbers of DWs. With the highest disorder strengths of 30 and 35 \%, the unambiguous 360$^\circ$ DWs are reached only in the late stage of the runs if at all. Animations of the magnetization reversal processes for different disorder strengths are provided as Supplemental Material~\cite{SM}. We ruled out the possibility of 360$^\circ$ DWs forming due to periodic boundary conditions by a similar calculation with open boundaries as presented in Fig.~S4.

Thus, stronger disorder induces the formation of a higher density of 360$^\circ$ DWs that also appear less stable -- presumably disorder lowers the energy barrier to remove a 360$^\circ$ DW, thus rendering the topological protection less effective. For example, in the case of 30 \% disorder, all the hysteresis curves collapse to $-1$ before $B_\textrm{ext}$ has reached $–25$ mT. Before collapsing, the DW width $w$ behaves like $w\sim |B_\mathrm{ext}|^{-1/2}$ as demonstrated in Fig.~\ref{fig:dwwidths}. The inverse square root relation between $w$ and the uniaxial anisotropy constant is a well-known result \cite{Chikazumi1997} for stationary 180$^{\circ}$ DWs, and our numerical examination verifies that $B_\textrm{ext}$ affects the immobile 360$^{\circ}$ DWs in a similar way as the uniaxial magnetocrystalline anisotropy affects 180$^{\circ}$ DWs. The disorder strength does not seem to affect the $w(|B_\textrm{ext}|)$ relation.


\section{Conclusions}
To conclude, our full micromagnetic simulations suggest that Barkhausen noise in Py thin films obeys neither predictions of nucleation nor front propagation models, but is instead a consequence of gradual formation of immobile 360$^{\circ}$ DWs via a sequence of abrupt magnetization rotation events. The criticality exhibited by the system appears to be disorder-induced such that the cutoff avalanche size is disorder-dependent. However, the "infinite avalanche" typically observed in Ising-type models for weak disorder is absent here. Instead, we observe smooth hysteresis curves for weak disorder, with progressively larger Barkhausen jumps as disorder is made stronger so that the critical disorder strength is approached from below. Disorder also controls the morphology of the ensuing 360$^{\circ}$ DWs such that for weak disorder almost straight DWs spanning the system are formed. Upon increasing the disorder strength, the DWs become increasingly rough, and finally form an irregular DW structure without clear, spanning 360$^{\circ}$ DWs at the critical disorder strength. The property of the Py samples not having any magnetocrystalline anisotropy results in the magnetization reversal process happening in a sequence of gradual and localized rotations of the magnetization, highlighting the importance of using full micromagnetic simulations to properly capture the details of the magnetization reversal process~\cite{Herranen2019}, and calling for experimental verification of our results, e.g., using magneto-optical imaging~\cite{schafer2007investigation}.

\begin{acknowledgments}
The authors acknowledge the computational resources provided by CSC, and support of the Academy of Finland via the Academy Project BarFume (Project no. 338955).
\end{acknowledgments}

\bibliography{bibliography}

\end{document}


\author{Sami Kaappa and Lasse Laurson}
\affiliation{Computational Physics Laboratory, Tampere University, P.O. Box 692, FI-33014 Tampere, Finland}
\title{Supplemental Material for \\ ``Barkhausen noise from formation of
360$^{\circ}$ domain walls \\ in disordered permalloy thin films''}
\maketitle

\noindent Here we briefly complement the methodology and results of the main article.\\ 

\noindent {\bf{Full field description}} The complete description of the effective field $\vec H_\mathrm{eff}(\vec r)$ of the system is given by
\begin{equation}
    \vec H_\mathrm{eff}(\vec r) = \vec H_\mathrm{ext}(\vec r) + \vec H_\mathrm{ex}(\vec r) + \vec H_\mathrm{demag}(\vec r),
\end{equation}
where $\vec H_\mathrm{ext}(\vec r)$ is the external field
\begin{equation}
    \vec H_\mathrm{ext}(\vec r) = \frac{1}{\mu_0}\vec B_\mathrm{ext},
\end{equation}
$\mu_0$ is the vacuum permeability, $\vec B_\mathrm{ext}$ is the applied field in Teslas, $\vec H_\mathrm{ex}(\vec r)$ the field contribution from the (Heisenberg) exchange interaction
\begin{equation}
    \vec H_\mathrm{ex}(\vec r) = \frac{2A}{\mu_0M_s(\vec r)}\nabla^2 \vec m(\vec r),
\end{equation}
$A$ the exchange coupling constant, and $\vec H_\mathrm{demag}(\vec r)$ the demagnetizing (magnetic dipole) field 
\begin{multline}
    \vec H_\mathrm{demag}(\vec r) = -\frac{1}{4\pi}\Biggl[\int_V \frac{(\vec r - \vec r^\prime) M_s(\vec r^\prime)\nabla\cdot \vec m(\vec r^\prime)}{\left|\vec r - \vec r^\prime\right|^3} \mathrm d\vec r^\prime\\
    + \int_S \frac{(\vec r - \vec r^\prime) M_s(\vec r^\prime)\hat n(\vec r^\prime)\cdot \vec m(\vec r^\prime)}{\left|\vec r - \vec r^\prime\right|^3} \mathrm d\vec r^\prime\Biggr]
\end{multline}
where the first integral is over the volume and the second integral over the surface of the system, $\hat n$ being the unit surface normal vector. These formulations are implemented in mumax3 as given in Ref. \cite{Vansteenkiste2014}.\\

\noindent {\bf{Justification of quasistatic mode.}} We compare the quasistatic and dynamic runs by simulating the magnetic reversal process in similar systems with both modes. To reduce the computational cost of the dynamical simulations, we begin the runs by driving the system quasistatically with the field step $\left|\Delta B_\mathrm{ext}\right| = 20$ \textmu T until $m_x$ is below 0.99, and the run is continued using the dynamic mode (time-integration of the LLG equation).

\begin{figure}[t!]
    \centering
    \includegraphics[width=\columnwidth]{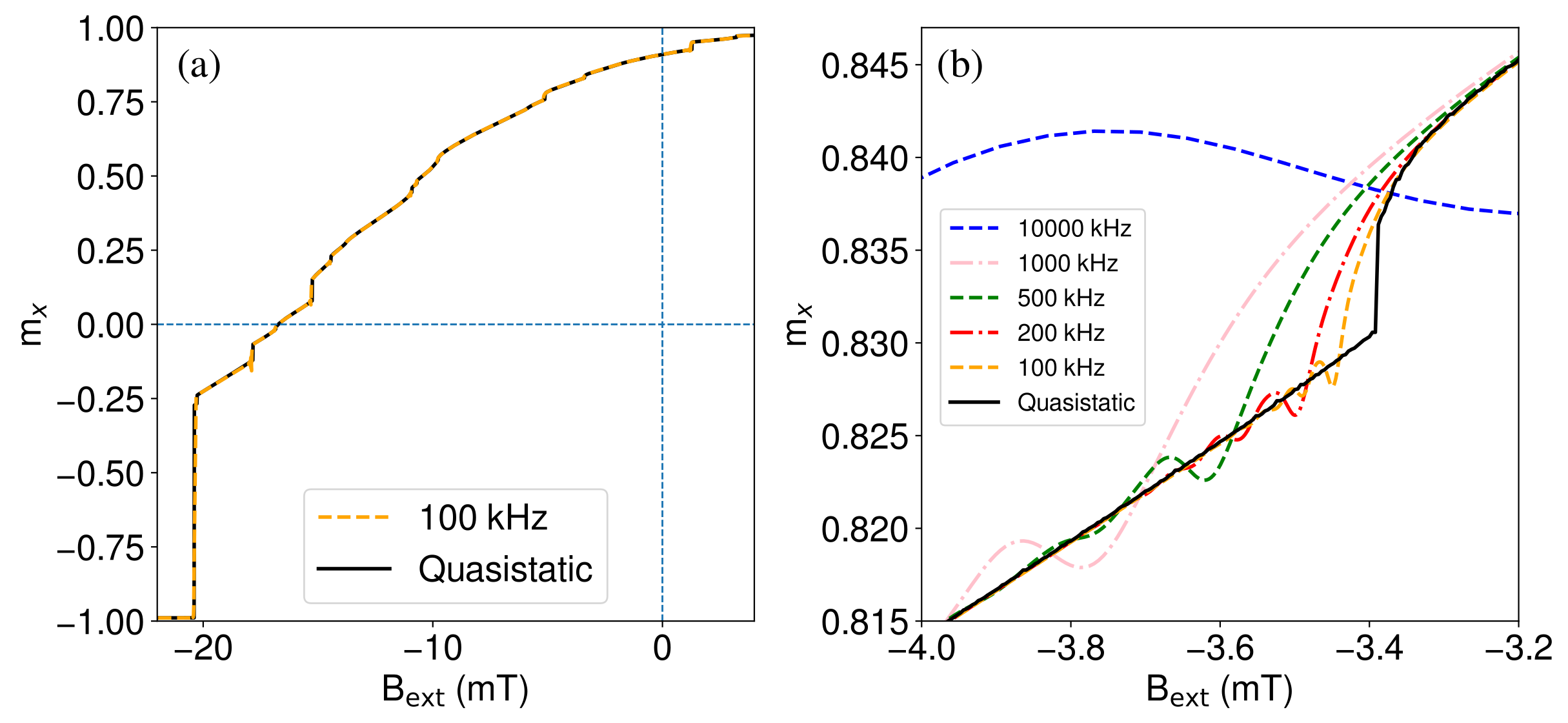}
    \caption{Comparison of the hysteresis curves between dynamic and quasistatic runs. The system size is 1024 nm $\times$ 1024 nm $\times$ 16 nm, and the disorder strength is 40 \%. (a) Halves of the hysteresis loops with quasistatic and dynamic modes. (b) A more detailed look at a single magnetization jump along the curve, showing the quasistatic result and the corresponding results from dynamic simulations with a wide range of field frequencies $f$. Notice how the dynamic curves gradually approach the quasistatic one as $f$ gets smaller.}
    \label{smfig1}
\end{figure}

Supplemental Fig.~\ref{smfig1}a shows halves of the hysteresis loops for simulations run in the quasistatic and dynamic modes. In the dynamic run, we use the RK45 algorithm \cite{RK45, Vansteenkiste2014} with an adaptive time step and apply the sinusoidal external field with frequency $f=100$ kHz. The Gilbert damping constant has the value of $\alpha=0.02$. In the quasistatic run, the field step $\left|\Delta B_\mathrm{ext}\right| = 4$ \textmu T is used. We see that the curves are qualitatively similar: the jumps are roughly the same size and they occur at roughly the same strength of the external field. This is a strong indication that the microscopic magnetization behaves locally the same way, that is the same magnetic structures and patterns are formed in the quasistatic and dynamical runs. We have confirmed that this is the case for the particular run by comparing the resulting magnetization configurations.

The details in the hysteresis curves are slightly different, however, as demonstrated in Supplemental Fig.~\ref{smfig1}b. The curves are shown for multiple frequencies in the dynamical runs. Focusing first on the 100 kHz curve, the most clear difference is that the Barkhausen jumps are abrupt in the quasistatic simulations, and they appear delayed in the dynamical curves due to the slowness of the response. After each jump, the dynamical curve experiences some vibration, which is then damped. The vibration is attributed to the damping precession of the local magnetizations, and its frequency is related to the material parameters in the LLG equation.

In Supplemental Fig.~\ref{smfig1}b, we further see that decreasing the frequency down from 10 MHz to 100 kHz leads to the dynamical curves approaching the quasistatic one. Based on this solid trend, we conclude that the intuitive hypothesis of the quasistatic mode corresponding to the limit $f\rightarrow 0$ is valid.

To further support the use of the quasistatic mode, let us discuss the computational requirements of the two modes. The system size of 1024 nm $\times$ 1024 nm $\times$ 16 nm and frequency of $f=100$ kHz, as used in the comparison above, is at the computational limit of the dynamical mode with dynamics governed by the RK45 algorithm: simulations of the dynamics of larger systems or smaller frequencies would lead to an unreasonably large requirement of computational time. However, for reliable Barkhausen jump statistics, we want to simulate larger system sizes to account for different emerging magnetic structures and their behaviour under changing applied field. On the other hand, the relevant frequency range in applications and experimental research is well below 1 kHz. Therefore, at this level of theory, we conclude that there are insurmountable challenges in modeling the Barkhausen noise statistics with dynamical simulations. We will run our main simulations with quasistatic mode, that can be used to mimic the dynamic runs with small frequencies. Also, larger system sizes can be simulated with the quasistatic mode, although a field step has to be selected that is small enough for reasonable accuracy and large enough for reasonable computational time.\\

\begin{figure}[t!]
    \centering
    \includegraphics[width=\columnwidth]{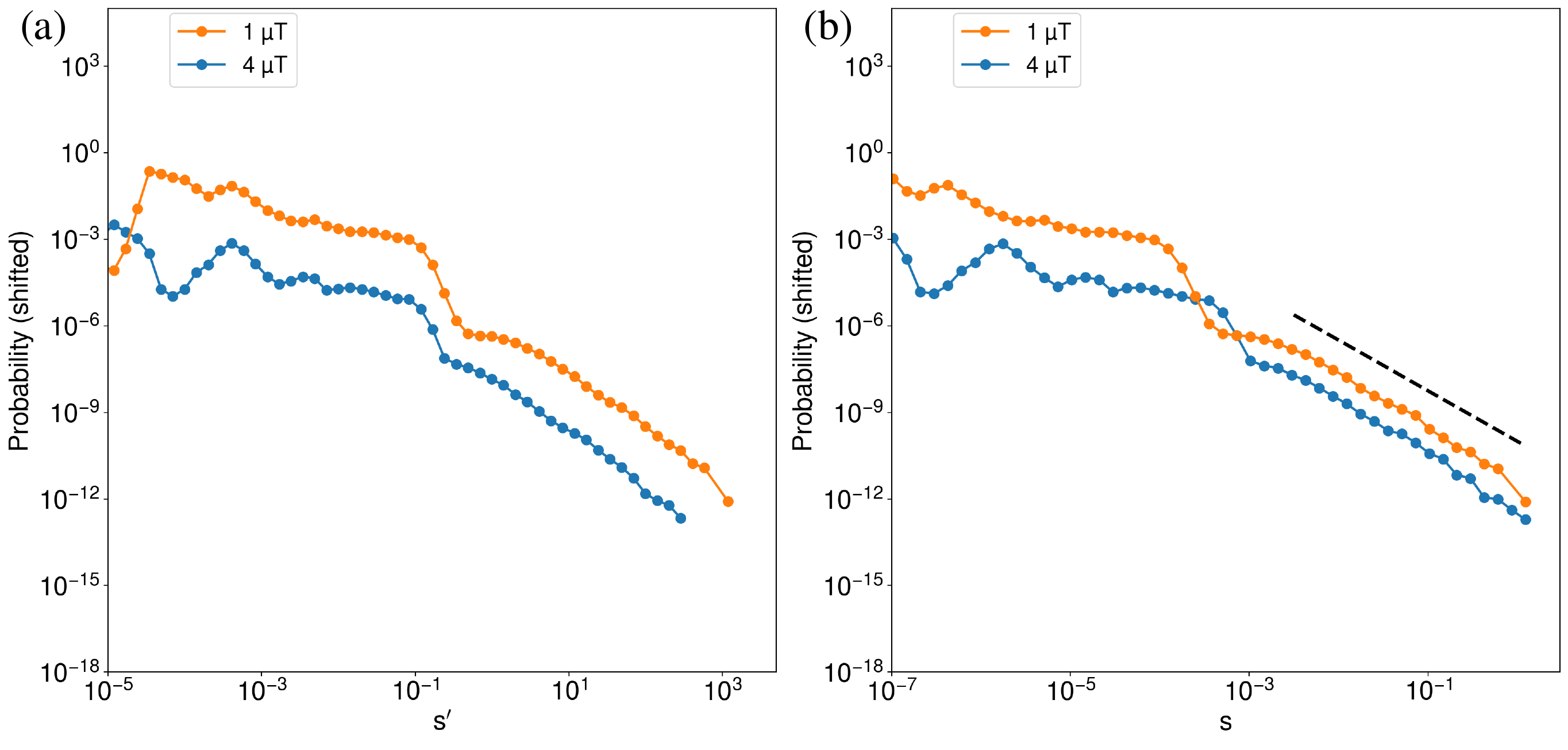}
    \caption{Comparison of the Barkhausen signal distributions with different field steps for disorder strength of 30 \%. (a) Plot against $s^\prime=\left|\Delta m_x/\Delta B_\mathrm{ext}\right|$ in units of mT$^{-1}$. (b) Plot against $s=\left|\Delta m_x\right|$. The magnitudes of the field steps are given in the legend. The dashed line indicates the fit power law with the exponent $\tau=1.75$ in accordance to the main text.}
    \label{smfig2}
\end{figure}

\noindent {\bf{Effect of the field step.}} In Supplemental Fig.~\ref{smfig2}, we show jump size distributions for the systems of 30 \% disorder, computed quasistatically with two different field step magnitudes. The distributions include 40 individual runs for both the field steps. The distributions are plotted both against $\left|\Delta m_x/\Delta B_\mathrm{ext}\right|$ and against $\left|\Delta m_x\right|$, where the latter corresponds to the figures in the main text. We see that the position of the bump is changed in the distribution for $\left|\Delta m_x\right|$, but not for the distribution for the quantity $\left|\Delta m_x/\Delta B_\mathrm{ext}\right|$. This supports the conclusion presented in the main text that the bump, as well as the part on its left hand side, corresponds to the smooth rotation of the magnetization. The part on the right hand side of the bump is preserved when the field step is changed, supporting the presence of the abrupt Barkhausen jumps in that size region.

There are no essential differences in the shape of the Barkhausen jump part of the distribution. Therefore, we conclude that the field step of 4 \textmu T that is used for the main results is sufficiently low for accurate results.\\

\noindent {\bf{Reversal processes with 30 \% disorder}} In Supplemental Fig.~\ref{smfig3}, different types of magnetization reversal processes in 30 \% disorder systems are illustrated along with the respective hysteresis curves. The selected runs represent the four subgroups that can be distinguished in the group of the hysteresis curves. We attribute the different subgroups to different densities of the forming 360$^\circ$ DWs. The large vertical jumps in the hysteresis curves correspond to vanishing of some of the DWs.

\begin{figure*}[t!]
    \centering
    \includegraphics[width=0.8\linewidth]{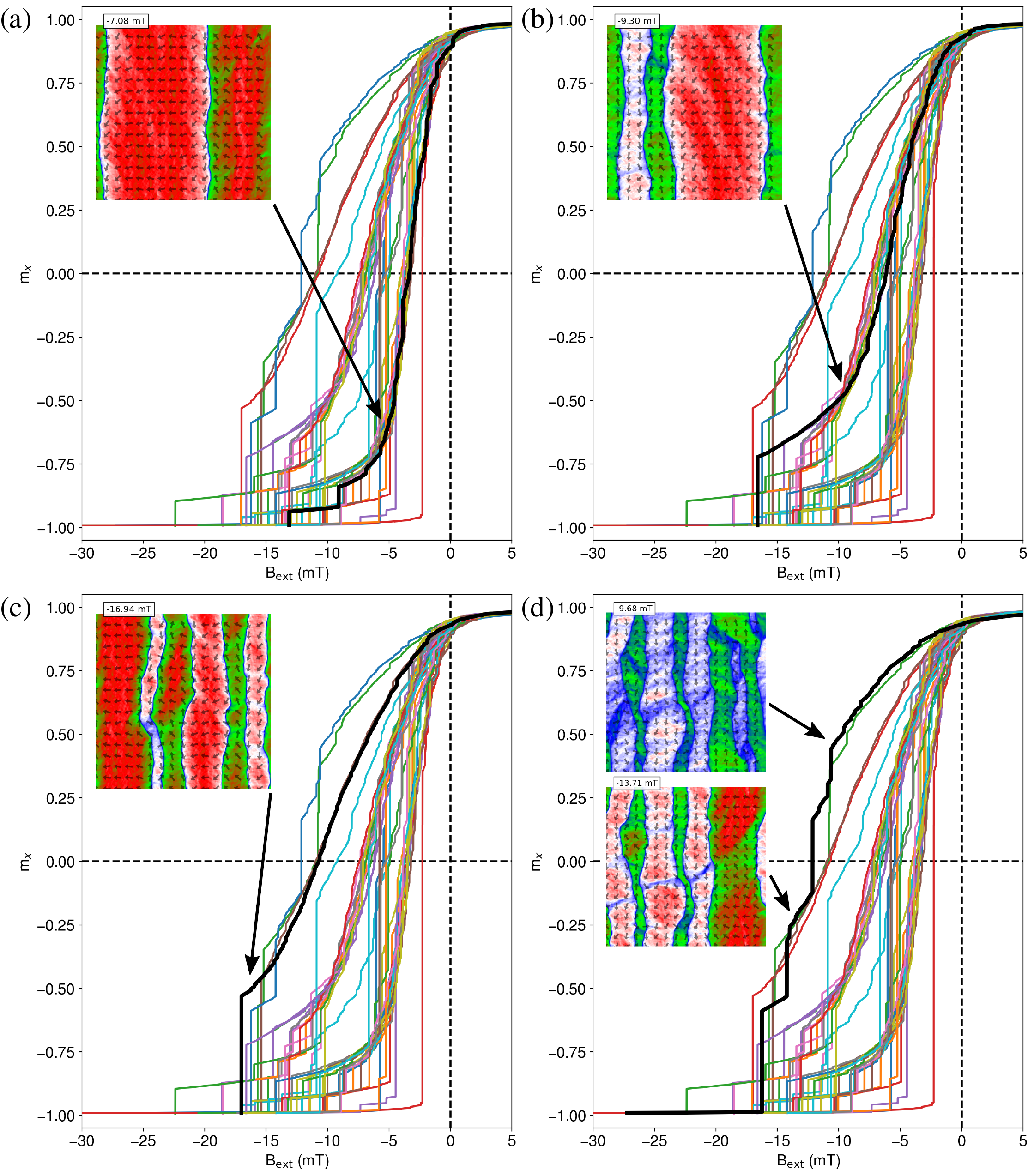}
    \caption{Magnetic configurations with 360$^\circ$ domain walls from selected runs with 30\% disorder strength. In the runs, (a) 2, (b) 4, (c) 6 distinct domain walls are formed. In (d), more walls are beginning to form, but some of them collapse during the run; at $-13.71$ mT, 6 domain walls can be distinguished as shown in the inset. The inset coloring corresponds to that of the main text. The respective hysteresis curves are drawn as thick, black lines.}
    \label{smfig3}
\end{figure*}

\begin{figure*}[t!]
    \centering
    \includegraphics[width=0.9\linewidth]{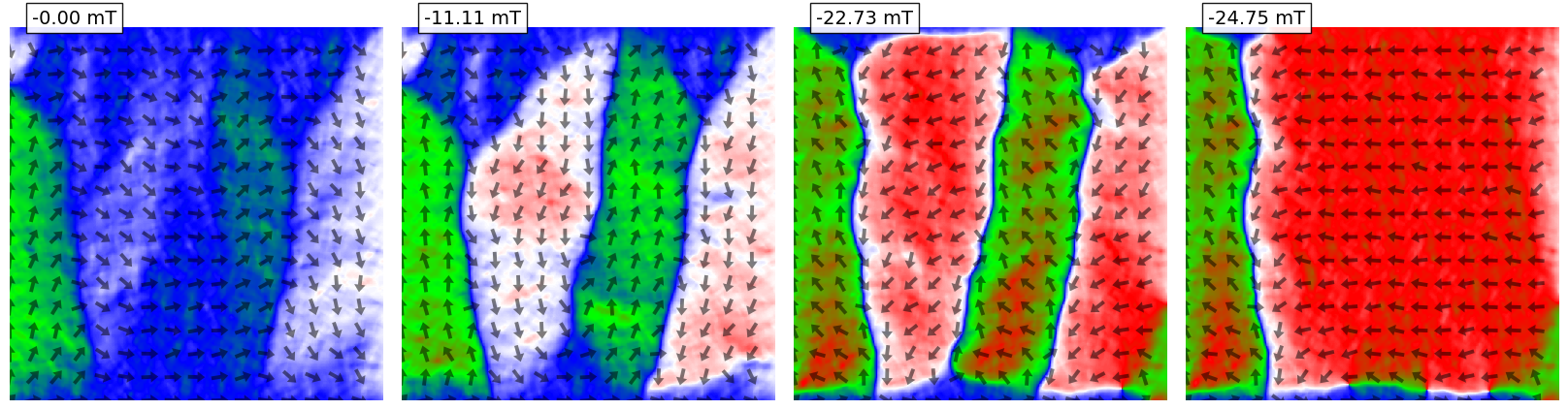}
    \caption{Magnetic configurations in a validation run with open boundary conditions. The disorder strength is 30 \%. The system size is 2 \textmu m $\times$ 2\textmu m with thickness of 16 nm. The field step is 0.1 \textmu T. Except for the formation of closure domain patterns near sample edges, similar dynamics of the magnetization to those in Fig. S3 are observed, validating that the formation of 360$^\circ$ DWs is not due to introduction of the periodic boundary conditions in the system.}
    \label{smfig4}
\end{figure*}

\bibliographystyle{apsrev4-2}
\bibliography{bibliography}